\documentclass[%
reprint,
twocolumn,
bibnotes,
amsmath,amssymb,
]{revtex4-2}
\usepackage{subcaption}
\usepackage{tikz}
\usepackage[font=footnotesize]{caption}
\usepackage{mwe}
\usepackage{graphicx}
\usepackage{dcolumn}
\usepackage{bm}
\usepackage{hyperref}

\usepackage{tikz,xcolor,hyperref}

\definecolor{lime}{HTML}{A6CE39}
\DeclareRobustCommand{\orcidicon}{%
	\begin{tikzpicture}
	\draw[lime, fill=lime] (0,0)
	circle [radius=0.16]
	node[white] {{\fontfamily{qag}\selectfont \tiny ID}};
	\draw[white, fill=white] (-0.0625,0.095)
	circle [radius=0.007];
	\end{tikzpicture}
	\hspace{-2mm}
}

\foreach \x in {A, ..., Z}{%
	\expandafter\xdef\csname orcid\x\endcsname{\noexpand\href{https://orcid.org/\csname orcidauthor\x\endcsname}{\noexpand\orcidicon}}
}

\begin{document}


\title{Spin-Hall conductivity and optical characteristics of noncentrosymmetric quantum spin Hall insulators: the case of PbBiI}

\author{Mohammad Mortezaei Nobahari\orcidA}
\affiliation{
Department of Physics, Ferdowsi University of Mashhad, Iran}

\author{Carmine Autieri\orcidA}
\affiliation{International Research Centre Magtop, Institute of Physics, Polish Academy of Sciences, Aleja Lotnik\'ow 32/46, 02668 Warsaw, Poland}




\date{\today}

\begin{abstract}
Quantum spin Hall insulators have attracted significant attention in recent years. Understanding the optical properties and spin Hall effect in these materials is crucial for technological advancements. In this study, we present theoretical analyses to explore the optical properties, Berry curvature and spin Hall conductivity of pristine and perturbed PbBiI using the linear combination of atomic orbitals and the Kubo formula. The system is not centrosymmetric and it is hosting at the same time Rashba spin-splitting and quantized spin Hall conductivity. Our calculations reveal that the electronic structure can be modified using staggered exchange fields and electric fields, leading to changes in the optical properties. Additionally, the spin Berry curvature and spin Hall conductivity are investigated as a function of the energy and temperature. The results indicate that due to the small dynamical spin Hall conductivity, generating an ac spin current in PbBiI requires the use of external magnetic fields or magnetic materials.
\end{abstract}

\maketitle


\section{\label{sec:1}Introduction }

In the realm of condensed matter physics, the emergence of topological materials has ushered in a new era of exploration, leading to the discovery of quantum phenomena with transformative implications. Among these materials, quantum spin Hall (QSH) insulators occupy a pivotal position, representing a paradigm shift in the understanding of topologically nontrivial electronic states~\cite{RevModPhys.82.3045}
The notion of a QSH insulator was first proposed by Bernevig $et$ $al.$~\cite{PhysRevLett.96.106802}, reflecting a revolutionary break from conventional electronic behavior by introducing the concept of topological protection for electronic states. These materials manifest insulating behavior in the bulk but host robust conducting edge states topologically protected against back-scattering by time-reversal symmetry, ushering in the promise of dissipationless electronic transport and novel spin-based functionalities~\cite{RevModPhys.83.1057,Hsieh2009,Bercioux_2015}.

Experimental investigations have validated the existence of QSH behavior in various material platforms, ranging from one- and two-dimensional systems to designed heterostructures, expanding the horizons of potential applications of these topological electronic states~\cite{doi:10.1126/science.1148047,Yan_2012,Brune2012,PhysRevLett.114.126802}. These experimental efforts have illuminated the intricate interplay between topological and electronic properties at the heart of QSH insulators.
Recent advances in experimental techniques, ranging from magneto-transport measurements to angle-resolved photoemission spectroscopy, have uncovered a plethora of materials showcasing QSH behavior, expanding the horizon of potential platforms for exploiting the remarkable attributes of these topological materials~\cite{PhysRevLett.61.2015,doi:10.1126/science.1234414}. Such strides in materials discovery and characterization open avenues for investigating the interplay between topological electronic states and intricate quantum phenomena.

Understanding the implications of QSH insulators extends beyond fundamental physics, venturing into the realm of practical applications in electronics and spintronics. The chiral nature of the edge states in QSH insulators offers the tantalizing prospect of dissipationless spin transport, holding promise for the development of efficient spin logic and memory devices that harness the spin degrees of freedom of electrons~\cite{Pesin2012,PhysRevB.82.045122,Jansen2012,Manchon2015}. Moreover, the intricate interplay between the topological and electronic properties of these materials underpins their potential for realizing topologically protected quantum computation and information processing~\cite{KITAEV20032,RevModPhys.80.1083}.
Recent theoretical advances have further underscored the potential of QSH insulators in redefining the limits of electronic and spin-based functionalities. The proposals for utilizing edge states in QSH insulators have opened up new avenues for achieving dissipationless spin transport and laying the groundwork for advancements in spin-based information processing and quantum computing~\cite{PhysRevLett.115.126803,PhysRevLett.102.256803}. The foundations set forth by the theoretical models have not only provided a roadmap for understanding the fundamental behavior of QSH insulators but also set the stage for exploring their transformative implications~\cite{PhysRevLett.106.106802,Zhang2009}. Amid these developments, the experimental realization of the quantum spin Hall effect and the identification of materials exhibiting topologically nontrivial electronic states have paved the way for exploring unique opportunities for harnessing their extraordinary properties~\cite{Xu2012,PhysRevLett.113.137201}. The ensuing synthesis of theory and experiment has propelled the field of topological electronics into a realm of unprecedented promise and potential.


The QSH insulating phase has been investigated in both centrosymmetric\cite{PhysRevB.76.045302,PhysRevB.106.245149,https://doi.org/10.1002/aelm.202300156} and noncentrosymmetric systems\cite{PhysRevLett.130.086202,PhysRevB.107.045138}, however, there are not so many cases where the QSH coexists with the Rashba spin splitting. In this paper, we will study a system where we have both the QSH effect and Rashba spin-splitting. 
Large Rashba spin-splitting is found in materials formed by heavy elements with strong intrinsic SOC
such as Bi, Pb, and W, among others~\cite{PhysRevLett.97.146803,PhysRevLett.104.066802,Yuan2013,PhysRevLett.101.266802}. To date, several types of QSHIs have been reported, and recently it proposed a honeycomb noncentrosymmetric QSHIs consisting of IV, V, and VII elements and Rashba-like SOC and unconventional spin texture.~\cite{PhysRevB.94.041302}
Until now, the properties of this material have been well studied in the presence of various disturbances. It has been shown that the thermodynamic properties of this material can be adjusted by a staggered exchange field~\cite{D2TA03132A}. Additionally, the effect of external fields on the electronic and optical properties of this material has also been well studied~\cite{10.1063/5.0097931,PhysRevB.106.165424,Phong_2021}

When there is no topological insulator phase, we cannot have the QSH phase but we can still have the ordinary spin Hall effect. The spin Hall conductivity (SHC) is a fundamental property of materials that describes the ability of a material to generate a spin current in response to an applied electric field.~\cite{kato_2004,PhysRevLett.94.047204,Shui,PhysRevLett.92.126603,Slawinska_2019,PhysRevMaterials.6.045004} This phenomenon arises from spin-orbit coupling, where the motion of electrons interacts with their spin degrees of freedom. In the presence of an electric field, electrons experience a transverse deflection due to the spin-orbit interaction, leading to the generation of a spin current perpendicular to the charge current. The SHC tensor quantifies this effect and provides valuable information about the spin dynamics in materials. Understanding and controlling the SHC is crucial for developing spintronic devices, such as spin-based transistors and memory storage devices, which rely on the manipulation of electron spins for information processing~\cite{Jungwirth2012,RevModPhys.76.323,HIROHATA2020166711}.

This paper begins by exploring the theoretical background in section \ref{sec:2} to gain insight into the properties of PbBiI. Next, theoretical frameworks are applied to calculate these properties in Section \ref{sec:3}, and the results are summarized in Section \ref{sec:4}.
                                 
\begin{figure}[]                                                   
\includegraphics[width=0.5\textwidth]{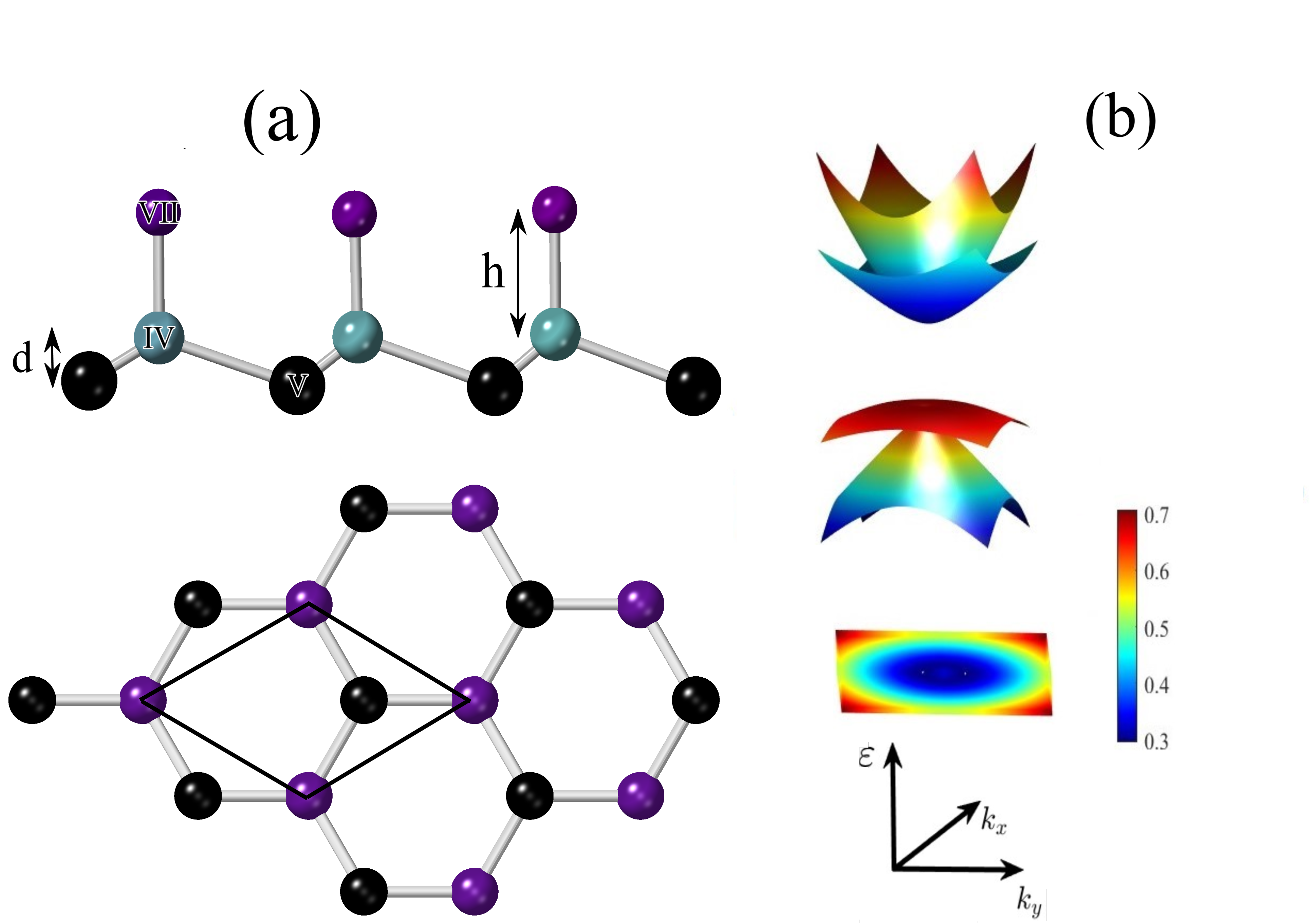}                
\caption{\label{fig:1} (a) Side and top view of the geometry structure of the PbBiI with Bi = V, Pb = IV, and I = VII by the buckled parameter $d \simeq 1.3$ and Pb-I $(h)$  
and Bi-Pb bond lengths 1.35 and 3.04 {\AA}, respectively. (b) 3D band structure and contour plot of $\varepsilon_3-\varepsilon_2$ in the $k_x$-$k_y$ plane.}             
\end{figure}                                                                                                                                                                                                                                                                                                                                                                                                                                                                                                             
\begin{figure}[]                                                                                                                                                          
\includegraphics[width=0.5\textwidth]{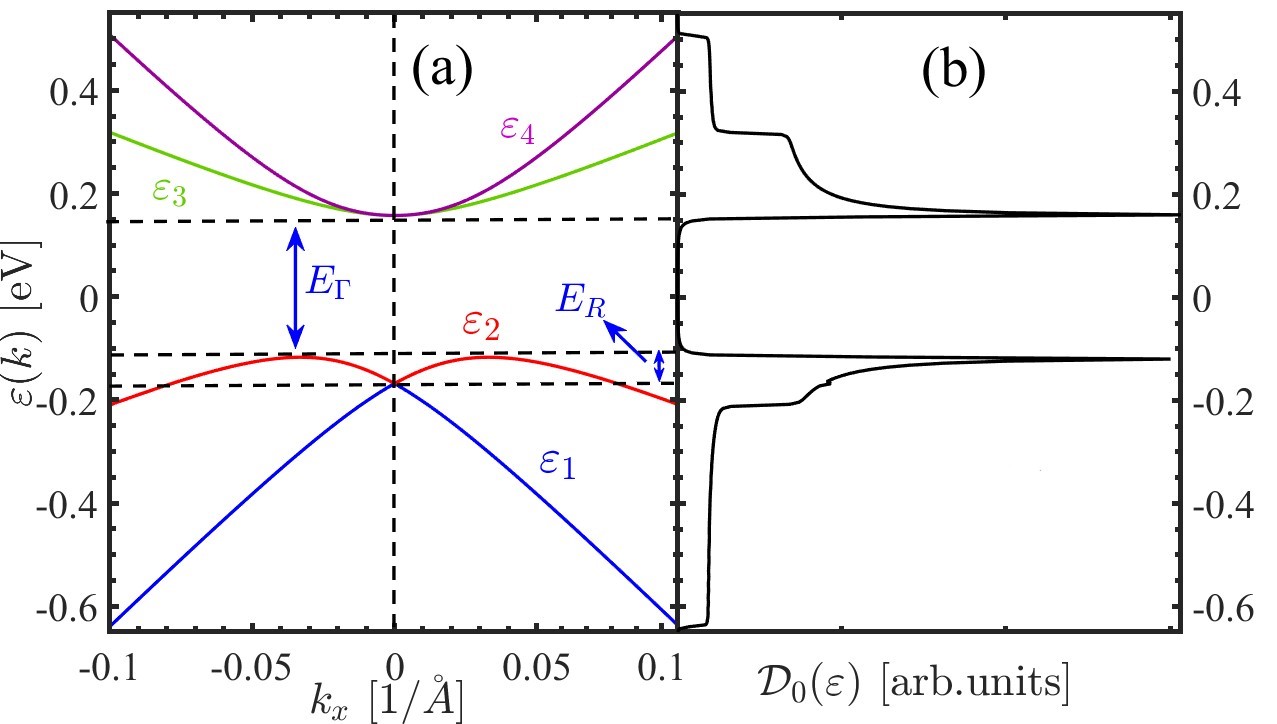}
\caption{\label{fig:2} (a) Band structure of the pristine PbBiI along the $k_x$ direction and $k_y=0$, and (b) the total density of states.}                                                           
\end{figure}

\section{Theory\label{sec:2}}

\subsection{Pristine and perturbed Hamiltonian}

The geometric structure of PbBiI is depicted in Fig. \ref{fig:1}(a) with top and side views, consisting of Bi (V), Pb (IV), and I (VII) elements. The distance parameters are approximately $d \approx$ 1.3 {\AA} and $h$ = 3.04 {\AA}. Previous analysis reveals that the highest valence comes from the $p_{x,y}$-Bi orbitals, while the $p_z$-Bi orbitals give the most relevant contribution to the lowest conduction band. As a result, we can ignore the Pb and I components in the electronic band structure of PbBiI. Hence, we focus on the single-particle bands with $l = 1$ (p-orbitals), $s = 1/2$ for spin angular momentum, and $j_{1,2} = \{1/2, 3/2\}$. The bands with j$_z$=$\pm$3/2 are far from the Fermi level~\cite{PhysRevB.94.041302}, and we have left the bands for two spin directions with $j_z = \pm1/2$. Therefore, the effective Hamiltonian in the basis of $|$j,j$_z\rangle$=$|\{1/2,3/2\}$,$\pm$1/2$\rangle$ can be expressed as:

\begin{align}
\mathcal H(\vec{k})&=\left({\begin{array}{cccc} -\varepsilon_{1/2}& 0&0&0\\
0 & -\varepsilon_{1/2} &0 & 0\\ 0 & 0&+\varepsilon_{3/2} & 0\\
0& 0 & 0 & +\varepsilon_{3/2}\\\end{array}}\right) \nonumber\\
&+\left({\begin{array}{cccc} \zeta_{1/2}k^2 & i \alpha_{R, 1/2}k_-&0&\gamma k_-\\
-i\alpha_{R, 1/2}k_+&\zeta_{1/2}k^2 & \gamma k_+ & 0\\ 0 & \gamma k_- & -\zeta_{3/2}k^2 & 0\\
\gamma k_+ & 0 & 0 &-\zeta_{3/2}k^2\\ 
\end{array}}\right)\label{eq:1},
\end{align}
The onsite energies are determined to be $\varepsilon_{1/2} = 0.1685$ eV and $\varepsilon_{3/2} = 0.1575$ eV while other parameters are obtained from $ab$ $initio$ calculations~\cite{PhysRevB.94.041302} and are $\zeta_{1/2} = 0.008187$ eV/{\AA$^2$}, $\zeta_{3/2} = 0.038068$ eV/{\AA}$^2$, $\alpha_{R, 1/2} = 3.0919$ eV/\AA, $\gamma = -3.5853$ eV/\AA, where $k_{\pm} = k_x \pm ik_y$, and $k = \sqrt{k_x^2 + k_y^2}$. The parameter $\alpha_{R, 1/2}$ represents the Rashba splitting in the conduction band, while $\gamma$ is the spin-orbit coupling between the valence and conduction band.

To introduce perturbations on the PbBiI system, external electric and magnetic exchange fields are applied to the Hamiltonian. The magnetic proximity effect arises from the induction of magnetic exchange fields in a material when it is in proximity to a ferromagnetic or antiferromagnetic substrate. These induced fields influence the orbital angular momentum within the basis, resulting in modifications to the Hamiltonian. Additionally, an external electric field can be applied by placing the PbBiI between two voltage gates. The modified Hamiltonian with perturbation terms $\mathcal H' (\vec{k})$ is expressed as:

\begin{equation}
\mathcal H' (\vec{k}) = \mathcal H (\vec{k}) + \mathcal H_{\mathcal R} + \mathcal H_{\mathcal I} \label{eq:2}\end{equation}
where $\mathcal H_{\mathcal R}$ and $\mathcal H_{\mathcal I}$ are the external staggered exchange field and electric field contributions respectively and are given by

\begin{equation}
\mathcal H_{\mathcal R}=-|j_z|\left({\begin{array}{cc} \mathcal R_{j_1}& 0\\
0 & \mathcal R_{j_2}\\\end{array}}\right)\otimes\sigma_z. \label{eq:3}
\end{equation}
and 
\begin{equation}
\mathcal H_{\mathcal I} = \left({\begin{array}{cccc} -\mathcal I/2& 0&0&0\\
0 & +\mathcal I/2 &0 & 0\\ 0 & 0&-\mathcal I/2 & 0\\
0& 0 & 0 & +\mathcal I/2\\\end{array}}\right)\label{eq:4}
\end{equation}

The induced exchange field $\mathcal R_{j_i}$ corresponds to the total angular momentum $j_i$ ($i = 1, 2)$~\cite{PhysRevLett.125.157402}, Here, $\sigma_z$ represents the $z$-component of the 3 $\times$ 3 Pauli matrix, and $\mathcal I$ can be controlled via electric field.

In Fig~\ref{fig:2} (a), the band structure of the unperturbed PbBiI system, obtained from Eq.~(\ref{eq:1}), is illustrated. This band structure comprises two valence and two conduction bands, where the valence band at the $\Gamma$ point is characterized by the states \{$|j = 3/2,  j_z\rangle$\}, and the effective state for the conduction band is \{$|j=1/2, j_z\rangle$\}. Consequently, the states include \{$|1/2, 1/2\rangle$\}, \{$|1/2, -1/2\rangle$\}, \{$|3/2, 1/2\rangle$\}, \{$|3/2, -1/2\rangle$\}.

\subsection{Density of states}
By utilizing the Green's function approach, the density of states (DOS) for the PbBiI can be computed. The DOS can be determined by adding up over the first Brillouin zone, 
	\begin{equation}
	D(\varepsilon)=-\frac{1}{N_c\pi}\sum_k{\Im[TrG(\vec{k},\varepsilon)]}\label{eq:5},
	\end{equation}
	where $N_c$ indicates the number of atoms in each unit cell. The non-interacting Green's function matrix is acquired through $G(\vec{k},\varepsilon)=[\varepsilon+i\eta-H(\vec{k})]^{-1}$, where $\eta$ represents the broadening factor  
	\begin{equation}
	G(\vec{k},\varepsilon)= \left({\begin{array}{cccc}G_{11} & G_{12} & G_{13} & G_{14}\\
		G_{21} & G_{22} & G_{23} & G_{24}\\
		G_{31} & G_{32} & G_{33} & G_{34}\\
		G_{41} & G_{42} & G_{43} & G_{44}\\
		\end{array}}\right)\label{eq:6},
	\end{equation}
	Using Eqs.~(\ref{eq:5}) and ~(\ref{eq:6}), the total DOS reads

	\begin{equation}
	\mathcal D_0(\varepsilon)=-\frac{1}{N_c\pi}\sum_k{\Im[G_{11}+G_{22}+G_{33}+G_{44}]}\label{eq:7}.
	\end{equation}

\subsection{Optical properties}
The optical conductivity tensor, $\sigma(\omega)$, can be determined using Ohm's law, which states that $J=\sigma E$, where $J$ is the current density, $E$ is the electric field, and $\sigma$ is the optical conductivity tensor. 
	
	\begin{equation}
		\sigma=\left({\begin{array}{cc} \sigma_{xx} & \sigma_{xy}\\
				\sigma_{yx}&\sigma_{yy}
		\end{array}}\right)\label{eq:8},
	\end{equation}
	
To calculate $\sigma(\omega)$, direction-dependent velocities are required. The current operator definition along the $\nu$ direction is $\overline{j}_{\nu}=e\partial \mathcal H'/\partial k_{\mu}$ 
	
	\begin{equation}
	\overline{j}_{\nu}=\left({\begin{array}{cccc}\zeta_{1/2}\frac{\partial k^2}{\partial k_{\nu}} & i\alpha_{R, 1/2}\frac{\partial k_-}{\partial k_{\nu}} & 0 & \gamma\frac{\partial k_-}{\partial k_{\nu}}\\
				- i\alpha_{R, 1/2}\frac{\partial k_+}{\partial k_{\nu}}&\zeta_{1/2}\frac{\partial k^2}{\partial k_{\nu}} & \gamma\frac{\partial k_+}{\partial k_{\nu}} & 0\\
				0 & \gamma\frac{\partial k_-}{\partial k_{\nu}} & -\zeta_{3/2}\frac{\partial k^2}{\partial k_{\nu}} & 0\\
				 \gamma\frac{\partial k_+}{\partial k_{\nu}} & 0 & 0 & -\zeta_{3/2}\frac{\partial k^2}{\partial k_{\nu}}
		\end{array}}\right)\label{eq:9},
	\end{equation}

	Also the general form of the current operator is\cite{PhysRevB.97.245408}

	\begin{equation}
		\overline{j}_{\nu}=-\frac{e}{\hbar}\sum_k{c_k^{\dag}c_k\alpha_k^{\nu}}+i\frac{e}{\hbar}\sum_k{c_k^{\dag}c_k\beta_k^{\nu}}\label{eq:10},
	\end{equation}

	that $\alpha_k^{\nu}$ and $\beta_k^{\nu}$ are intraband and inter-band direction-depended velocities along the $\nu$-direction.
	
	By using linear response theory, the optical conductivity is given as
	
	\begin{equation}
		\sigma_{\nu \nu'}(\omega)=\frac{g_s}{\hbar\omega S}\int{dt e^{i\omega t}\langle[\overline{j}_{\nu}(t),\overline{j}_{\nu'}(0)]\rangle}\label{eq:11},
	\end{equation}
	
	where $g_s=2$ is the spin degeneracy, $\omega$ is photon frequency and $S$ is the 2D planar area.\par
	
	Using Eq.~(\ref{eq:11}), the  interband optical conductivity is given as~\cite{Yarmohammadi_2020,PhysRevB.97.245408,Mortezaei_Nobahari2023-vc}:
	
	\begin{equation}
	\begin{split}	
			\sigma_{\nu\nu'}^{inter}(\omega)&=-i\frac{g_se^2}{\hbar^2S}\sum_{k}\sum_{j,j'}\sum_{j_zj_z'}{\beta_{j_zj_z'}^{jj' \nu}(\vec{k})\beta_{j_zj_z'}^{jj'\nu'}(\vec{k})}\\
			&\times\frac{1}{\varepsilon_{\vec{k},j_z}^{j}-\varepsilon_{\vec{k},j_z'}^{j'}}\frac{f_{\vec{k},j_z}^{j}-f_{\vec{k},j_z'}^{j'}}{\hbar\omega-\varepsilon_{\vec{k},j_z}^{j}+\varepsilon_{\vec{k},j_z'}^{j'} +i\eta_1}\label{eq:12}.
			\end{split}
	\end{equation}
where $f_{\vec{k}, j_z}^{j} = 1/(1+\exp{((\varepsilon_{\vec{k}, j_z}^{j} - \mu_0)/k_BT))}$ is the Fermi-Dirac distribution at a constant temperature $T$ and chemical potential $\mu = 0$, $\varepsilon_{\vec{k}, j_z}^{j}$ represents the eigenvalue of the energy, $\eta_1$ denotes the finite damping between the conduction and valence bands, and $\beta_{j_zj_z'}^{jj' \nu}(\vec{k})=\langle \vec{k}; j, j_z |j_{\nu}| \vec{k}; j', j_z'\rangle$ and $\beta_{j_zj_z'}^{jj' \nu'}(\vec{k})=\langle \vec{k};j', j_z' |j_{\nu'}| \vec{k}; j, j_z\rangle$ are velocities along the $\nu$ and $\nu'$-directions respectively.\par

Another important optical property is the electron energy loss spectroscopy (EELS). The energy electron loss spectrum is a type of spectroscopy technique used to study the electronic properties of materials. It involves measuring the energy lost by electrons as they interact with a sample, which can provide information about the electronic structure and bonding of the material. The spectrum is generated by bombarding the sample with high-energy electrons and then measuring the energy distribution of the scattered electrons. The resulting spectrum can reveal details about the valence and conduction bands of the material, as well as the presence of impurities or defects. To calculate EELS we need the dielectric function which is given by:

\begin{equation}
\varepsilon^{\nu\nu'}(\omega)-\varepsilon_r=\frac{i\sigma_{\nu\nu'}^{inter}(\omega)}{\omega\varepsilon_0d_{BP}}\label{eq:13},
\end{equation}

where $\varepsilon_r$ is the relative permittivity and $d_{BP}$ is the PbBiI thickness. One can calculate the EELS as 

\begin{equation}
L_{\nu\nu'}(\omega)=-\Im[\frac{1}{\varepsilon^{\nu\nu'}(\omega)}]=\frac{\varepsilon_2^{\nu\nu'}(\omega)}{(\varepsilon_1^{\nu\nu'}(\omega))^2+(\varepsilon_2^{\nu\nu'}(\omega))^2}\label{eq:14}
\end{equation}

We can determine the reflectivity by using the refractive index $n$ and extinction coefficient $\kappa$ and dielectric function. We have

\begin{equation}
n_{\nu\nu}(\omega)=\frac{1}{\sqrt{2}}\sqrt{|\varepsilon^{\nu\nu}(\omega)|+\varepsilon_1^{\nu\nu'}(\omega)}\label{eq:15}
\end{equation}

and

\begin{equation}
\kappa_{\nu\nu'}(\omega)=\frac{1}{\sqrt{2}}\sqrt{|\varepsilon^{\nu\nu'}(\omega)|-\varepsilon_1^{\nu\nu'}(\omega)}\label{eq:16}
\end{equation}

that we have write $\varepsilon_{\nu\nu'}=\varepsilon_1^{\nu\nu'}+i\varepsilon_2^{\nu\nu'}$. Reflectivity can be calculated as
\begin{equation}
R_{\nu\nu'}(\omega)=\frac{(1-n_{\nu\nu'}(\omega))^2+\kappa_{\nu\nu'}^2(\omega)}{(1+n_{\nu\nu'}(\omega))^2+\kappa_{\nu\nu'}^2(\omega)}\label{eq:17}
\end{equation}

\subsection{Spin Hall conductivity}

We calculate both static ($\omega=0$) and dynamic ($\omega\neq 0$) SHC using the Kubo formula and Berry curvatures. The component $\sigma_{xy}^z$ of the SHC tensor            
represents a spin current flowing along the $x$-direction, polarized along the $z$ and an electric field applied along the $y$-axis. The Kubo formula for the SHC is\cite{PhysRevLett.100.096401,PhysRevLett.94.226601}:                                                                     
\begin{equation}	                                                                                                                                                        
\sigma^{SH}(\omega)=\frac{e}{\hbar}\sum_k\sum_{j,j_z}  f^J_{\vec{k},j_z}\Omega_{j',j'_z}^z \label{eq:18}	                                                                              
\end{equation} 
     
where dynamic spin Berry curvature, velocity, and spin-current operators are defined as 

 \begin{equation}                                                                                                                                                          
\Omega_{j',j'_z}^z(\vec{k},\omega)=\sum _{j', j'_z} \frac{\zeta_{j_z,j'_z,z}^{j,j', x}(\vec{k})\beta_{j_z,j'_z}^{j,j',                                                          
y}(\vec{k})}{(\varepsilon_{\vec{k},j_z}^{j}-\varepsilon_{\vec{k},j'_z}^{j'})^2-(\hbar\omega+i\eta)^2}.\label{eq:19}                                                                                       
\end{equation}                                                                                                                              
and the static spin Berry curvature definition is                                                                                                                        
\begin{equation}                                                              \Omega_{j',j'_z}^z(\vec{k})=\sum _{j', j'_z} 2\text{Im}\frac{\zeta_{j_z,j'_z,z}^{j,j', x}(\vec{k})\beta_{j_z,j'_z}^{j,j',                                                          
y}(\vec{k})}{(\varepsilon_{\vec{k},j_z}^{j}-\varepsilon_{\vec{k},j'_z}^{j'})^2}. \end{equation}\label{eq:20}                                                                                                                                                                                                                                                                                                                 
and 
                                                                                                                   
 \begin{equation}                                                                                                                                                         
\beta_{j_z,j_z'}^{j,j, y}(\vec{k})=\langle \vec{k}; j, j_z |j_{y}| \vec{k}; j', j_z'\rangle                                                                               
\end{equation}\label{eq:21}                                                                                                                                               
                                                                                                                                                                          
\begin{equation}                                                                                                                                                          
\zeta_{j_z,j_z, z}^{j,j, x}(\vec{k})=\langle \vec{k}; j, j_z |S_x^z| \vec{k}; j', j_z'\rangle                                                                             
\end{equation}\label{eq:22}                                                                                                                                               
where $S_x^z=\frac{\hbar}{4}\{\beta\Sigma_z , j_x\}$ while $\beta$ and $\Sigma_z$ are the $4\times4$ Dirac matrices~\cite{PhysRevLett.94.226601}.

\begin{figure}[]                                                                                                                                                          
\includegraphics[width=0.5\textwidth]{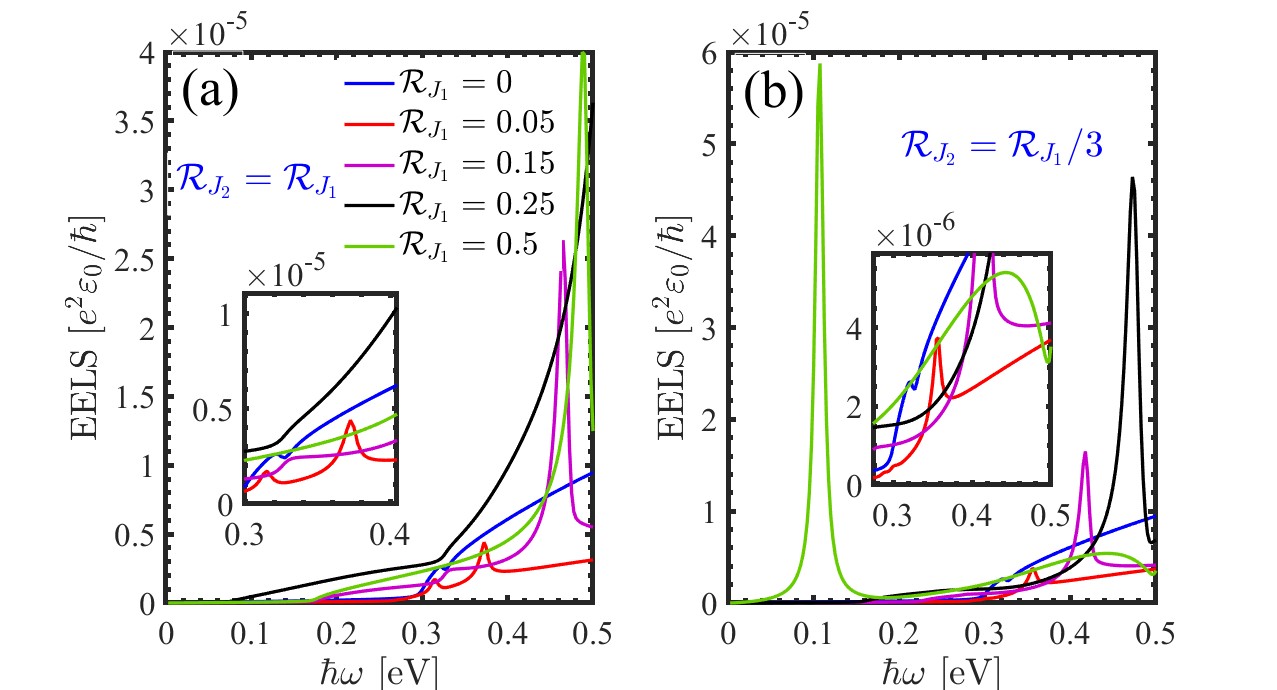}
                                                      
\caption{\label{fig:3} EELS obtained from Eq.~(\ref{eq:14} in presence of the staggered exchange field with (a) $\mathcal R_{j_2} =\mathcal R_{j_1}$ and (b) $\mathcal R_{j_1} = \mathcal R_{j_2}/3$}.                                
\end{figure}                                                                                                                                                              
                            
\begin{figure}                                                                                                                                                            
  \centering                                                                                                                                                              
  \includegraphics[width=0.5\textwidth]{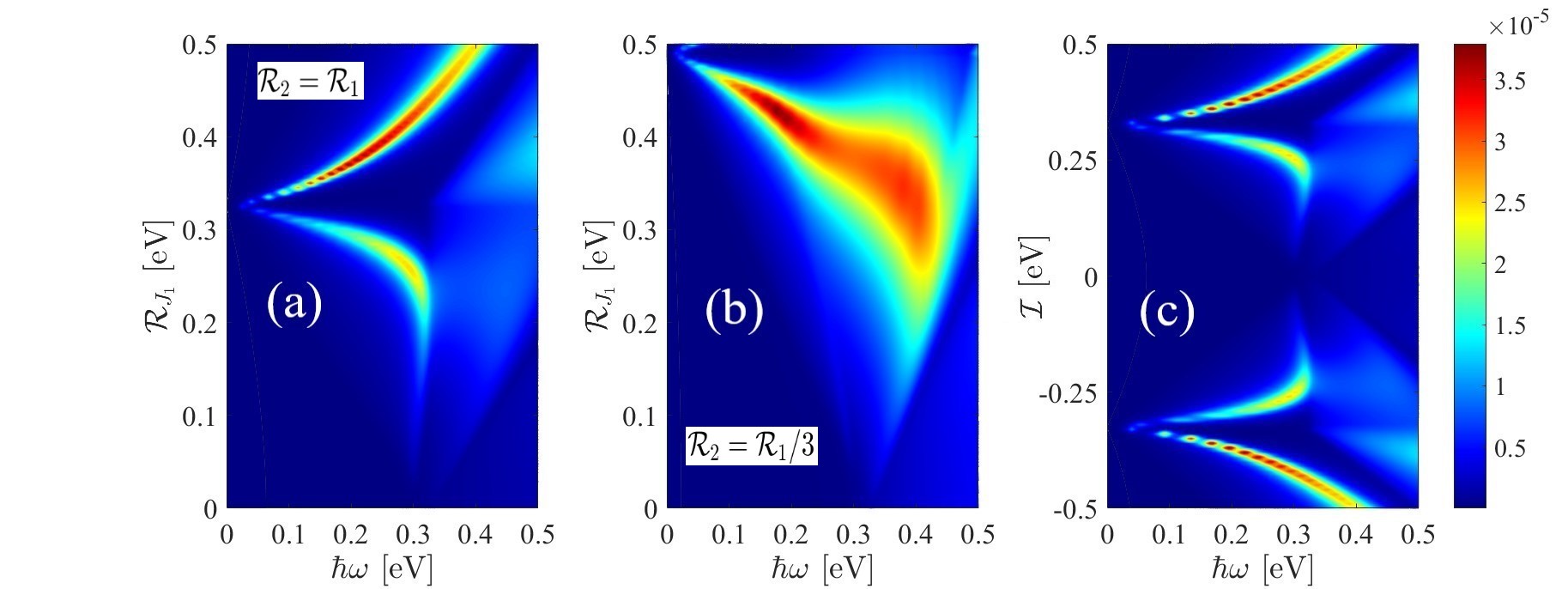}                                                                                                                        
  \caption{Color density of the EELS in the presence of (a) the staggered exchange field with $\mathcal R_{j_2} =\mathcal R_{j_1}$, (b) $\mathcal R_{j_2} =\mathcal R_{j_1}/3$ and (c) external electric field.}\label{fig:4}                                                                                                                                              
\end{figure}   

\begin{figure}[]                                                                                                                                                          
\includegraphics[width=0.5\textwidth]{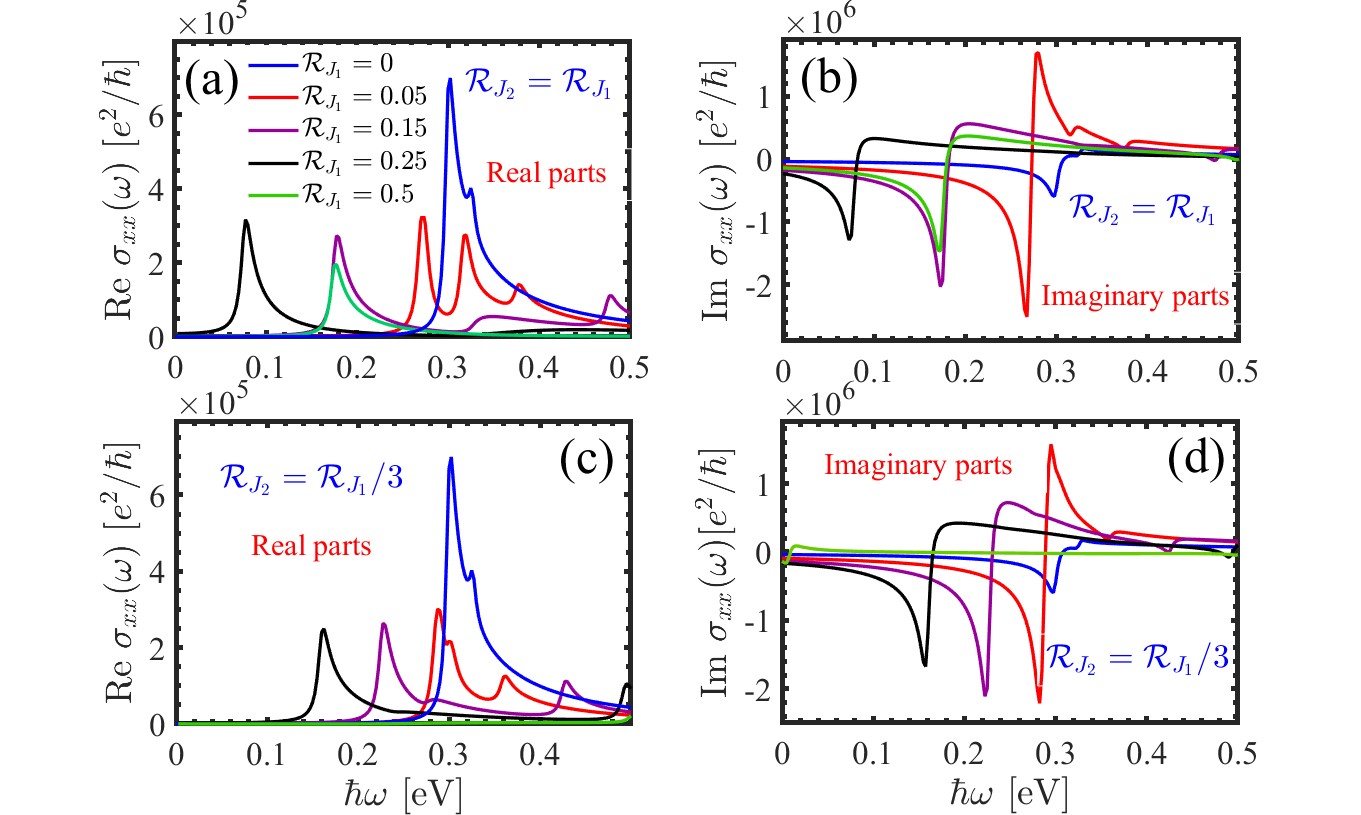}
\caption{\label{fig:5} Real and imaginary parts of the optical conductivity for polarized light along the $x$-axis by introducing staggered exchange field (a), (b) $\mathcal R_{j_2} =\mathcal R_{j_1}$ and (c), (d) $\mathcal R_{j_2} =\mathcal R_{j_1}/3$}                                                                                 
\end{figure}

\begin{figure}[]                                                                                                                                                          
\includegraphics[width=0.5\textwidth]{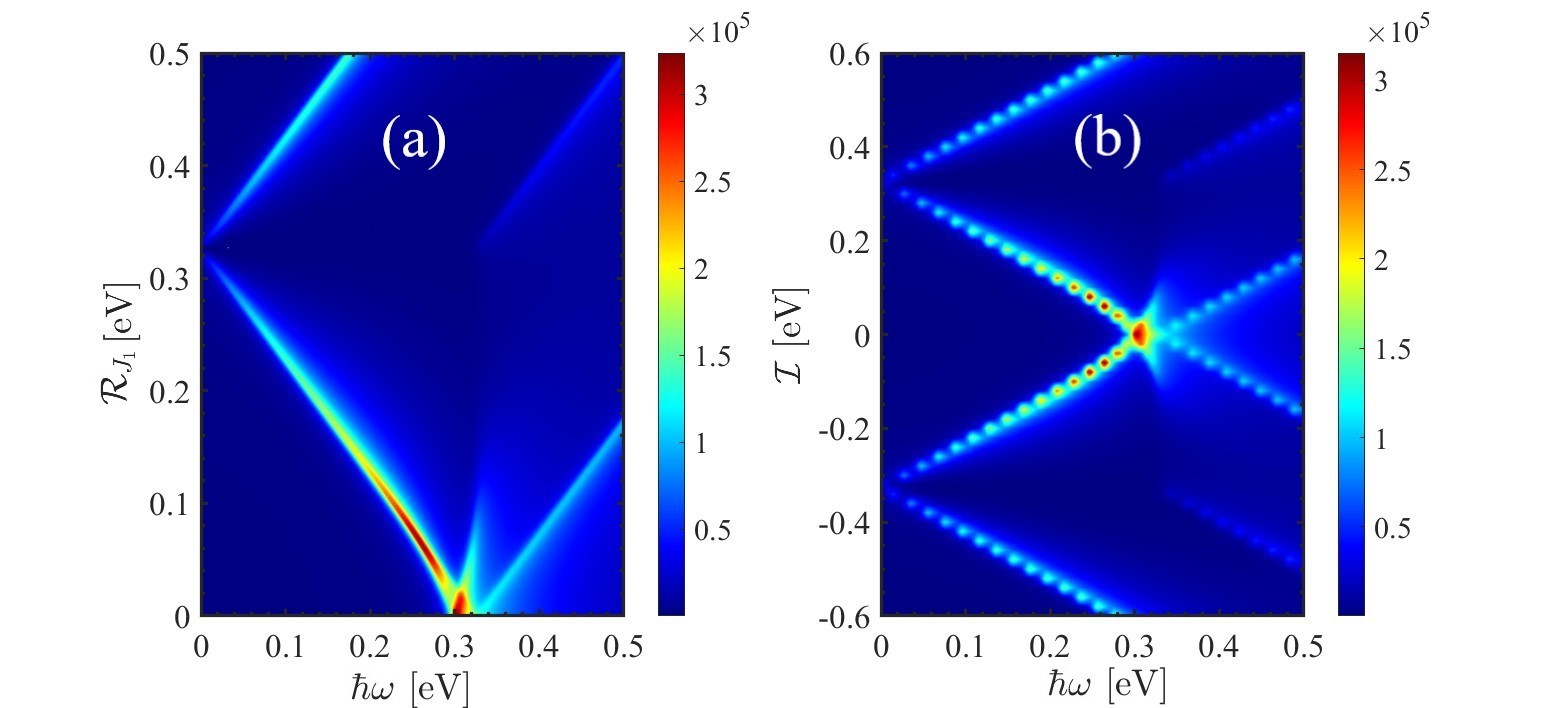}
\caption{\label{fig:6} Color density of the real part of the optical conductivity in subject to the external perturbation (a) staggered exchange field ( $\mathcal R_{j_2} =\mathcal R_{j_1}$) and (b) electric field.}                              
\end{figure}                                                                                                                                                              
                                                                                                                                                                          
\begin{figure}[]                                                                                                                                                          
\includegraphics[width=0.5\textwidth]{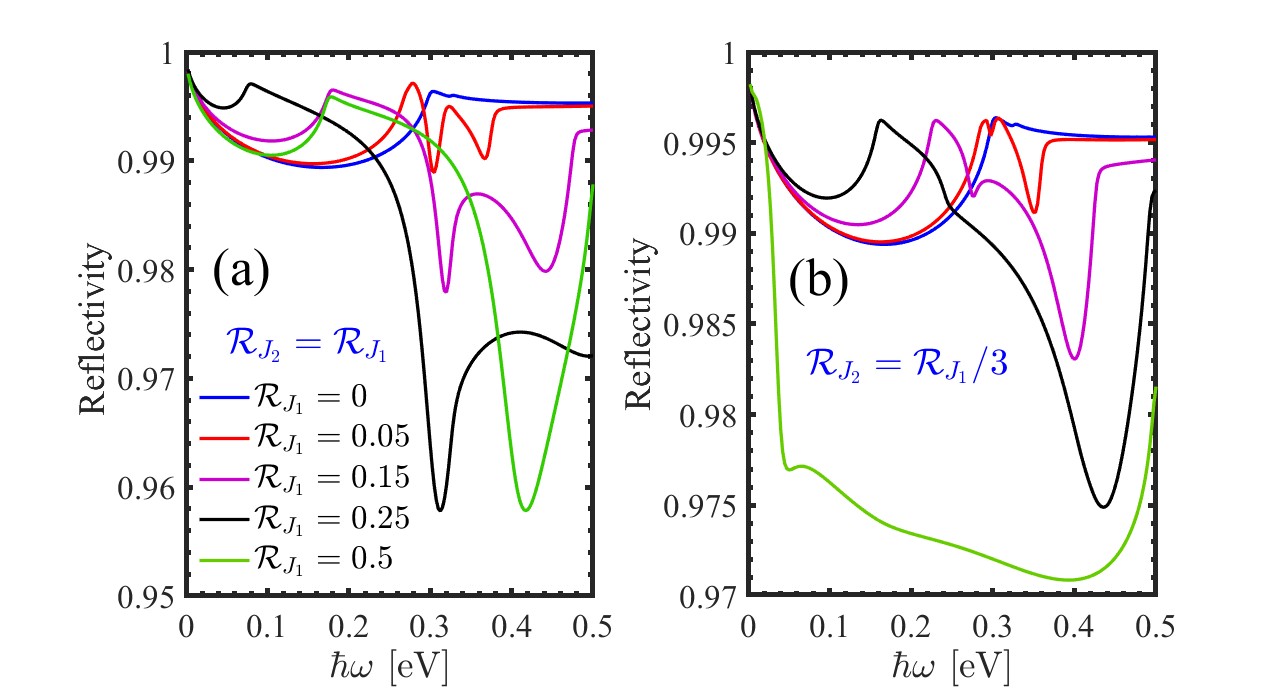}
\caption{\label{fig:7} Reflectivity of the perturbed PbBiI in the presence of staggered exchange field (a) $\mathcal R_{j_2} =\mathcal R_{j_1}$ and (b) $\mathcal R_{j_2} =\mathcal R_{j_1}/3$}                              
\end{figure}                                                                                                                                                              
                                                                                                                                                                          
\begin{figure}[]                                                                                                                                                          
\includegraphics[width=0.5\textwidth]{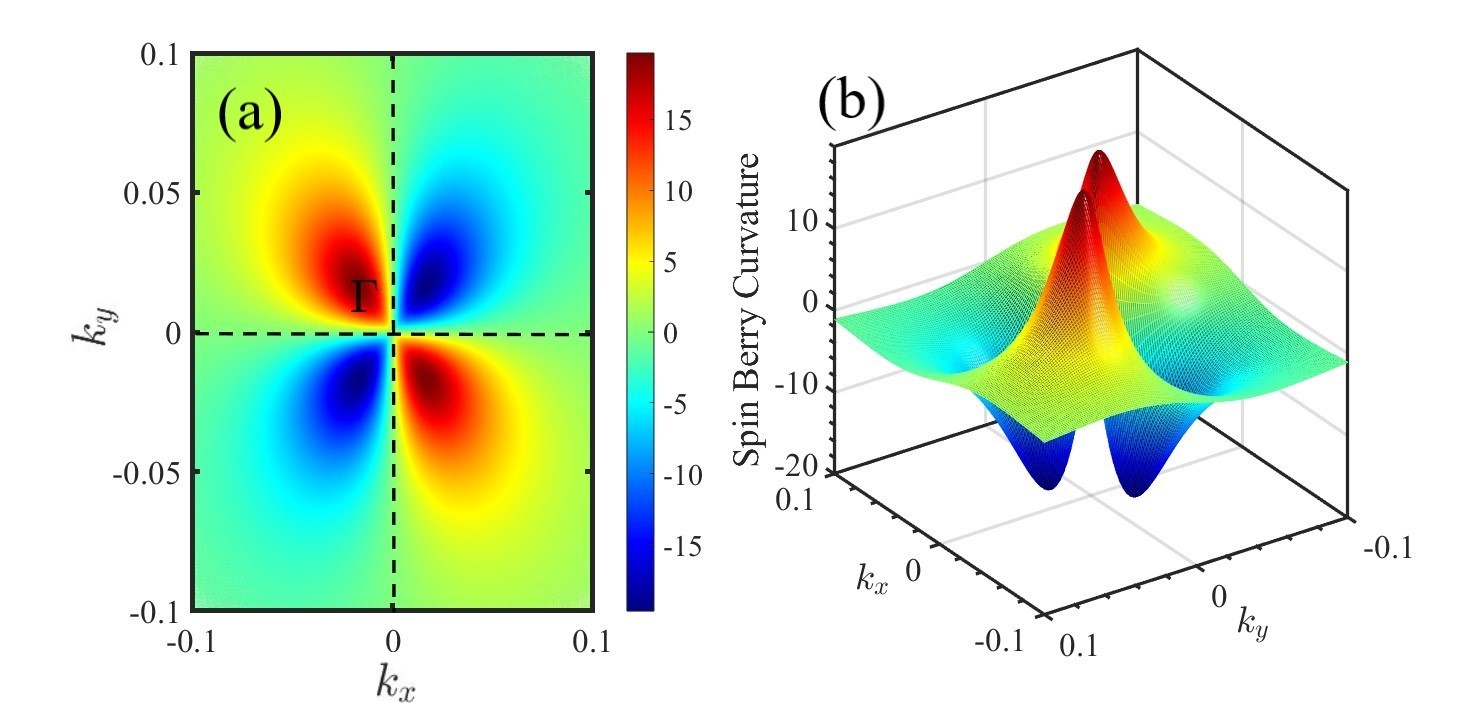}
\caption{\label{fig:8} Spin Berry curvature in the Brillouin zone around the $\Gamma$ point in the form of (a) color density and (b) surface plot.}                              
\end{figure}                                                                                                                                                              
                                                                                                                                                                          
\begin{figure}[]                                                                                                                                                          
\includegraphics[width=0.5\textwidth]{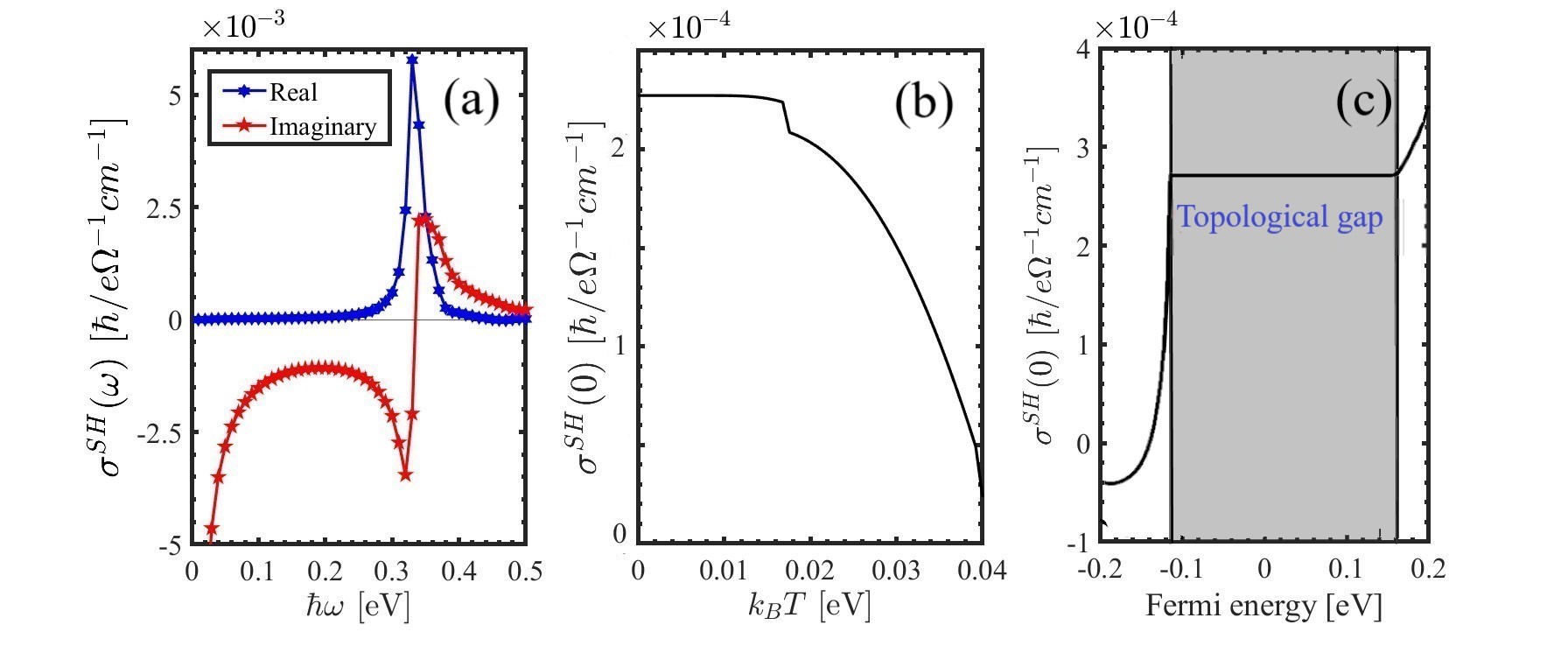}
\caption{\label{fig:9}Calculated (a) real and imaginary parts of the dynamical SHC, (b) statical SHC as a function of temperature, and (c) statical SHC versus Fermi energy.  }                              
\end{figure}                                                                                                                                                              
                       
\section{Results and Discussion\label{sec:3}}            

The main results of the paper are discussed in this Section.
Fig.~\ref{fig:3} displays the EELS results under the influence of a staggered exchange field for varying values of $\mathcal R_{j_1}$ and $\mathcal R_{j_2}$. The range considered for $\mathcal R_{j_1}$ is between 0 and 0.5 eV, while we have considered two ratios for  $\frac{\mathcal R_{j_2}}{\mathcal R_{j_1}}$=1 or 1/3. In the case where $\mathcal R_{j_2} = \mathcal R_{j_1}$ (as shown in Fig.~\ref{fig:3}(a)), distinct peaks are observed, and as the strength of the field increases, the peaks shift towards higher energies. Conversely, for $\mathcal R_{j_2} = \mathcal R_{j_1}/3$ (depicted in Fig.~\ref{fig:3} (b)), an opposite shift is observed for $\mathcal R_{j_1} = 0.5$ eV. To explore the entire                                                   
 spectrum of staggered and electric fields, contour plots of the EELS have been calculated within a specific energy and external field range (refer to                                                              
 Fig.~\ref{fig:4}). Notably, the majority of EELS behavior is associated with $\mathcal R_{j_2}=\mathcal R_{j_1}/3$ and $\mathcal R_{j_1}> 0.1$ eV. Comparing                                                       
 Fig.~\ref{fig:4} (a) and (c) reveals a similarity in the EELS response to positive values when both a staggered exchange field ($\mathcal R_{j_2} = \mathcal R_{j_1}$)                                             
 and an external electric field is applied.\par                                                                                                                                                                                                         
 The optical conductivity of PbBiI with external perturbations is computed using the Kubo formula. Due to PbBiI's isotropic nature, we focused on the optical conductivity along the $x$-axis and omitted the $y$-axis. In the pristine case, a peak in the real part of the optical conductivity aligns with the band gap energy (see Fig.~\ref{fig:5} (a) and (c)). Adjusting the $\mathcal R_{j_1}$ and $\mathcal R_{j_2}$ parameters alters the optical conductivity and shift peak energies. It is evident that regardless of the $\mathcal R_{j_1}$ and $\mathcal R_{j_2}$ ratio, introducing a staggered exchange field leads to new peaks in the real parts, with only their positions changing based on different ratios. Furthermore, due to the Kramers-Kronig relation, a dip in the imaginary parts occurs at the peak's energy in the real parts.\par                                                                                    
        
Figure~\ref{fig:6} showcases a contour plot illustrating the optical conductivity as a function of frequency, staggered exchange field (Fig.~\ref{fig:6} (a)), and electric field (Fig.~\ref{fig:6} (b)). It is evident from Fig.~\ref{fig:6} (a) that the peak of the optical conductivity appears at an energy of 0.3 eV. When $\mathcal R_{j_1}$=0.3 eV, it causes a shift towards lower energies however, for $\mathcal R_{j_1} >$ 0.3 eV this trend is reversed. The external electric field also has a similar effect except in negative magnitudes.\par                                                                                                                                                    
The reflectivity is defined as the ratio of the intensity of reflected light to the intensity of incident light, typically expressed as a percentage. It is a key parameter in numerous applications, including optics, coatings, architecture, and solar energy technologies, where controlling and optimizing the reflective properties of materials is essential for achieving desired performance characteristics. Fig.~\ref{fig:7} is related to the reflectivity in the presence of the staggered exchange field. As we can see by increasing the photon's frequency we have an increase in the reflectivity and the peaks appear. In addition by comparing Fig.~\ref{fig:7} (a) and (b) we found that reflectivity is greater in case $\mathcal R_{j_2} = \mathcal R_{j_1}$.\par                                                                                                                       
                                                                                                                                                                                                                    
The 3D plot and color density of the spin Berry curvature of the PbBiI in the Brillouin zone are shown in Fig.~\ref{fig:8} (a) and (b) respectively. The Berry curvature is enhanced in locations where the energy difference between the bands gets reduced as in the anticrossing points. According to the figures, the Berry curvature is maximum around the $\Gamma$ point and decreases when moving away from this point. This is due to the existence of the band crossing near the $\Gamma$ point.

The SHC can be expressed in terms of the spin Berry curvature (see Eq.~(20)). Fig.~\ref{fig:9} (a) represents the dynamical SHC of the PbBiI versus frequency. Both the real and imaginary parts of the ac SHC are small. This suggests that to generate an ac spin current, one needs to use a magnetic field or magnetic materials.

The dependence of the dc SHC on temperature is illustrated in Fig.~\ref{fig:9} (b), however, a critical temperature of $k_BT = 0.016$ eV is identified, beyond which the SHC decreases sharply towards room temperature, reaching a minimum for $k_BT > 0.025$ eV. The Spin Hall conductivity of the PbBiI as a function of Fermi energy is plotted in Fig.~\ref{fig:9} (c). The SHC has a quantized value within the topological band gap. Our calculations reveal that the SHC is minimal at $E_F=0$ and maximum at $E_F=±0.2$ eV. This is because there are band crossings induced by spin-orbit interactions at these specific energies.

                                                                                                                                                                          





\section{Conclusions\label{sec:4}}
In summary, we have investigated the noncentrosymmetric system PbBiI where quantized spin Hall conductivity and Rashba spin-splitting coexist. Our analysis involved the computation of the Berry curvature and spin Hall conductivity, along with investigating the electronic and optical characteristics under external influences. By introducing staggered exchange and electric fields, we were able to manipulate the optical conductivity and EELS of the PbBiI. The peak of the real part of the optical conductivity is observed at 0.3 eV, with perturbations causing a shift towards lower energies. The Berry curvature reaches its maximum near the $\Gamma$ point where band crossing occurs, diminishing significantly further away from this region. Given the low dynamical spin Hall conductivity, a magnetic field is necessary to induce an a.c. spin current. Furthermore, the dc spin Hall conductivity exhibits critical behavior around $k_BT = 0.016$ eV.

\begin{acknowledgments}
C. A. was supported by the Foundation for Polish Science project "MagTop" no. FENG.02.01-IP.05-0028/23 co-financed by the European Union from the funds of Priority 2 of the European Funds for a Smart Economy Program 2021–2027 (FENG).
\end{acknowledgments}

\section*{Data availability}
The data that support the findings of this study are available from the corresponding author upon reasonable request.
\bibliography{bib}
\end{document}